\documentclass[prd,preprint,superscriptaddress,showpacs,byrevtex]{revtex4}
\usepackage{bm}
\def\fsl#1{\setbox0=\hbox{$#1$}           
   \dimen0=\wd0                                 
   \setbox1=\hbox{/} \dimen1=\wd1               
   \ifdim\dimen0>\dimen1                        
      \rlap{\hbox to \dimen0{\hfil/\hfil}}      
      #1                                        
   \else                                        
      \rlap{\hbox to \dimen1{\hfil$#1$\hfil}}   
      /                                         
   \fi}                                         %
\newcommand{\be}{\begin{equation}}
\newcommand{\ee}{\end{equation}}
\newcommand{\bea}{\begin{eqnarray}}
\newcommand{\eea}{\end{eqnarray}}
\newcommand{\beq}{\begin{equation}}
\newcommand{\eeq}{\end{equation}}
\newcommand{\beqs}{\begin{eqnarray}}
\newcommand{\eeqs}{\end{eqnarray}}

\newcommand{\aslash}{A\hspace{-0.067in}\slash}

\begin{document}
\title{ Hadron Formation From Non-Equilibrium Quark-Gluon Plasma at RHIC and LHC Using Closed-Time Path Integral Formalism }
\author{Gouranga C Nayak }\thanks{E-Mail: nayakg138@gmail.com}
%
%
\date{\today}
\begin{abstract}
Recently we have reported the correct formulation of the hadron formation from the quarks and gluons by using the lattice QCD method at the zero temperature. Similarly we have also reported the correct formulation of the hadron formation from the thermalized quark-gluon plasma by using the lattice QCD method at the finite temperature. In this paper we extend this to non-equilibrium QCD and present the correct formulation of the hadron formation from the non-equilibrium quark-gluon plasma by using the closed time path integral formalism. Hadron formation from the non-equilibrium quark-gluon plasma is necessary to detect the quark-gluon plasma at RHIC and LHC.
\end{abstract}
\maketitle
\pagestyle{plain}

\pagenumbering{arabic}

\section{introduction}

There has been lot of efforts to produce the quark-gluon plasma (QGP) in the laboratory. The QGP existed in the early universe just after $10^{-12}$ seconds of the big bang. Hence it is important to recreate this early universe scenario in the laboratory \cite{pnd,pnd1,pnd2}. The temperature of the quark-gluon plasma is $\ge$ 200 MeV ($\ge 10^{12}$ Kelvin) which corresponds to the energy density $\ge$ 2~GeV/fm$^3$ which is higher than the energy density ($\sim$ 0.15~GeV/fm$^3$) of the normal nucleus.

There are two experiments which study the quark-gluon plasma formation in the laboratory: 1) relativistic heavy-ion colliders (RHIC) at BNL and 2) large hadron colliders (LHC) at CERN. The RHIC collides the heavy nuclei Au-Au at the total center of mass energy $\sqrt{s}$ = 39.4 TeV and the LHC collides the heavy nuclei Pb-Pb at the total center of mass energy $\sqrt{s}$ = 1044.16 TeV. Since the total center of mass energies at these heavy-ion colliders are very high which are deposited over the small volumes just after the nuclear collisions, there is no doubt that the initial energy densities created in these nuclear collisions are much higher than the required energy density ($\sim$ 2 GeV/fm$^3$) to produce the quark-gluon plasma.

However, the major challenge is to detect the quark-gluon plasma at RHIC and LHC and to study its properties. This is because we have not directly experimentally observed quarks and gluons. In QCD in vacuum we measure hadronic cross section at the high energy colliders which is calculated from the partonic scattering cross section by using the factorization theorem in QCD \cite{fcd,fcd1,fcd2}. Hence we cannot directly detect the quark-gluon plasma at RHIC and LHC. For this reason indirect signatures are proposed for its detection. The major indirect signatures for the quark-gluon plasma detection at RHIC and LHC are: 1) $j/\psi$ suppression, 2) jet quenching, 3) strangeness enhancement, 4) dilepton production and 5) diphoton emission.

Note that since the two nuclei at RHIC and LHC travel almost at the speed of light the quark-gluon plasma formed at RHIC and LHC may be in non-equilibrium from the beginning to the end. This is because the hadronization time scale in QCD is $\sim$ 1 fm/c. This time is too small for many more secondary partonic collisions to take place to bring the system into equilibrium.

If the quark-gluon plasma is in non-equilibrium at RHIC and LHC from the beginning to the end then it becomes extremely difficult to detect the quark-gluon plasma by using the indirect signatures. This is because all the indirect hadronic signatures have to calculated in non-equilibrium and non-perturbative QCD by using the closed-time path integral formalism in non-equilibrium QCD which is an extremely difficult problem.

Note that although there have been lot of progress in the perturbative QCD (pQCD) but there has not been much progress in the non-perturbative QCD. In the renormalized QCD \cite{tvd} the coupling increases as the distance increases due to asymptotic freedom \cite{gwd,pod}. Since the hadron formation from quarks and gluons is a long distance phenomenon in QCD one needs to solve the non-perturbative QCD to study the hadron formation from the quarks and gluons. However, it is impossible to solve the non-perturbative QCD analytically due to the presence of cubic and quartic gluonic field terms in the QCD lagrangian in the path integration in QCD [see section II for details].

For this reason the lattice QCD is used to numerically compute the path integration in QCD in the Euclidean time to study the hadron formation from the quarks and gluons in QCD in vacuum. Since the Euclidean time can be related to the inverse of the temperature, the lattice QCD is also applicable at the finite temperature QCD to study the thermalized quark-gluon plasma.

Recently we have reported the correct formulation of the hadron formation from the quarks and gluons by using the lattice QCD method at the zero temperature \cite{nkcd} by incorporating the non-zero boundary surface term in QCD which arises in the energy conservation equation in the Noether's theorem in QCD \cite{nkntd} due to the confinement of quarks and gluons inside the finite size hadron \cite{nkffd}. Similarly we have also reported the correct formulation of the hadron formation from the thermalized quark-gluon plasma by using the lattice QCD method at the finite temperature \cite{nkftd} by incorporating this non-zero boundary surface term in QCD. We have also studied the parton to hadron fragmentation function using the lattice QCD method at zero temperature in \cite{nkfg0}. Similarly we have studied the  	
parton to hadron fragmentation function from the quark-gluon plasma using the lattice QCD method at the finite temperature in \cite{nkfgt}.

However, as explained above the actual quark-gluon plasma at RHIC and LHC may be in non-equilibrium from the beginning to the end. Hence it is necessary to study the hadron formation from the non-equilibrium quark-gluon plasma in order to detect the quark-gluon plasma at RHIC and LHC.

In this paper we present the correct formulation of the hadron formation from the non-equilibrium quark-gluon plasma by using the closed time path integral formalism in non-equilibrium QCD. Hadron formation from the non-equilibrium quark-gluon plasma is necessary to detect the quark-gluon plasma at RHIC and LHC.

The paper is organized as follows. In section II we discuss the hadron formation from the quarks and gluons by using the lattice QCD method at the zero temperature. In section III we discuss the hadron formation from the thermalized quark-gluon plasma by using the lattice QCD method at finite temperature. In section IV we present the correct formulation of the hadron formation from the non-equilibrium quark-gluon plasma at RHIC and LHC by using the closed-time path integral formalism in non-equilibrium QCD. Section V contains conclusions.

\section{ Hadron formation from quarks and gluons using lattice QCD method at zero temperature }

The partonic operator ${\cal Q}^H(x)$ to study the hadron $H$ is chosen in such a way that it contains the same quantum numbers of the hadron $H$. For example, for the pion $\pi^+$ formation the partonic operator is given by
\bea
{\cal Q}^{\pi^+}(x) ={\psi^\dagger}^d(x)\gamma_5 \psi^u(x)
\label{ppd}
\eea
where $\psi^u(x)$ is the up quark field and $\psi^d(x)$ is the down quark field. Similarly, for the proton $P$ formation the partonic operator is given by
\bea
{\cal Q}^P(x) =\psi^u(x)\psi^d(x)C\gamma_5 \psi^u(x)
\label{prd}
\eea
where $C$ is the charge conjugation operator and for the neutron $N$ formation the partonic operator is given by
\bea
{\cal Q}^N(x) =\psi^d(x)\psi^u(x)C\gamma_5 \psi^d(x).
\label{nud}
\eea
In the path integral formulation the non-perturbative correlation function of the partonic operator ${\cal Q}^H(x)$ to study the hadron $H$ formation in QCD at zero temperature is given by
\bea
&&<0| {\cal Q}^H(x') {\cal Q}^H(0)|0> = \frac{1}{Z[0]} \int [dA] [d{\bar \psi}] [d\psi] \times {\cal Q}^H(x') {\cal Q}^H(0) \times {\rm det}[\frac{\delta J_f^a}{\delta \omega^b}] \nonumber \\
&& \times {\rm exp}[i\int d^4x [-\frac{1}{4} F_{\nu \lambda}^c(x)F^{\nu \lambda c}(x) -\frac{1}{2\alpha} [J_f^c(x)]^2 +{\bar \psi}_k(x)[\delta^{ki}(i{\not \partial}-m)+gT^c_{ki}\aslash^c(x)]\psi_i(x)]] \nonumber \\
\label{pid}
\eea
where $|0>$ is the non-parturbative vacuum state (not the pQCD vacuum state but the full QCD vacuum state), $J_f^a(x)$ is the gauge fixing term, $\alpha$ is the gauge fixing parameter, $\psi_i$ is the quark field, $A_\mu^a(x)$ is the gluon field, $m$ is the quark mass and
\bea
F_{\nu \lambda}^c(x) =\partial_\nu A_\lambda^c(x)-\partial_\lambda A_\nu^c(x) +gf^{chs} A_\nu^h(x) A_\lambda^s(x).
\label{fnld}
\eea
Note that in eq. (\ref{pid}) we do not have any ghost fields because we directly work with the ghost determinant ${\rm det}[\frac{\delta J_f^a}{\delta \omega^b}]$ in this paper. A typical choice of the gauge fixing term in the covariant gauge is given by $J_f^c(x) =\partial^\nu A_\nu^c(x)$.

In the Heisenberg representation the time evolution of the partonic operator is given by
\bea
{\cal Q}^H(t,r)=e^{-itH^{QCD}_{q{\bar q}g}}{\cal Q}^H(0,r) e^{itH^{QCD}_{q{\bar q}g}}
\label{ted}
\eea
where $H^{QCD}_{q{\bar q}g}$ is the QCD Hamiltonian of all the quarks plus antiquarks plus gluons inside the hadron $H$. Using eq. (\ref{ted}) in (\ref{pid}) we find
\bea
&&<0| {\cal Q}^H(t',r') {\cal Q}^H(0)|0> = <0| e^{-it'H^{QCD}_{q{\bar q}g}}{\cal Q}^H(0,r') e^{it'H^{QCD}_{q{\bar q}g}} {\cal Q}^H(0)|0>.
\label{pi1d}
\eea
Using
\bea
H^{QCD}_{q{\bar q}g}|0> =0
\label{zrd}
\eea
in eq. (\ref{pi1d}) we find
\bea
&&<0| {\cal Q}^H(t',r') {\cal Q}^H(0)|0> = <0|{\cal Q}^H(0,r') e^{it'H^{QCD}_{q{\bar q}g}} {\cal Q}^H(0)|0>.
\label{pi2d}
\eea
Inserting a complete set of hadronic energy eigenstates
\bea
\sum_l |H_l><H_l|=1
\label{csd}
\eea
in eq. (\ref{pi2d}) we find
\bea
&&<0| {\cal Q}^H(t',r') {\cal Q}^H(0)|0> = \sum_l <0|{\cal Q}^H(0,r')|H_l><H_l| {\cal Q}^H(0)|0>e^{i\int dt'E^{QCD}_{q{\bar q}g,~l}(t')}
\label{pi3d}
\eea
where $E^{QCD}_{q{\bar q}g,~l}(t)$ is the energy of all the quarks plus antiquarks plus gluons inside the hadron $H$ in its l$th$ level and $\int dt'$ is an indefinite integration.

Note that $e^{i\int dtE^{QCD}_{q{\bar q}g,~l}(t)}$ is oscillatory in the Minkowski time (real time) and hence one can not neglect the higher energy level contribution of the hadron $H$. For this reason one goes to the Euclidean time (imaginary time) $\tau$ defined by
\bea
-it=\tau
\label{imgd}
\eea
to find from eq. (\ref{pi3d})
\bea
&&<0| {\cal Q}^H(\tau',r') {\cal Q}^H(0)|0> = \sum_l <0|{\cal Q}^H(0,r')|H_l><H_l| {\cal Q}^H(0)|0>e^{-\int d\tau'E^{QCD}_{q{\bar q}g,~l}(\tau')}.\nonumber \\
\label{pi4d}
\eea
Since $e^{-\int d\tau E^{QCD}_{q{\bar q}g,~l}(\tau)}$ is exponentially decaying one can neglect the higher energy level contribution of the hadron $H$ in the large Euclidean time limit to find from eq. (\ref{pi4d})
\bea
&&\sum_{r'}[<0| {\cal Q}^H(\tau',r') {\cal Q}^H(0)|0>]_{\tau'\rightarrow \infty} = |<0|{\cal Q}^H(0)|H>|^2e^{-\int d\tau'E^{QCD}_{q{\bar q}g}(\tau')} \nonumber \\
\label{pi5d}
\eea
where $E^{QCD}_{q{\bar q}g}(t)$ is the energy of all the quarks plus antiquarks plus gluons inside the hadron $H$ in its ground state with
\bea
|H>=|H_{l=0}>
\label{gsd}
\eea
being the energy eigenstate of the hadron $H$ in its ground state.

The energy of the hadron $E^H$ is given by \cite{nkcd}
\bea
E^H=E^{QCD}_{q{\bar q}g}(t)+E^{QCD}_{\rm flux}(t)
\label{hed}
\eea
where $E^{QCD}_{\rm flux}(t)$ is the non-zero energy flux in the energy conservation equation in the Noether's theorem in QCD \cite{nkntd}. The energy flux $E^{QCD}_{\rm flux}(t)$ is non-zero due to the confinement of quarks and gluons inside the finite size hadron \cite{nkffd}. The non-zero energy flux in QCD can be calculated by using the lattice QCD method at the zero temperature by using the formula \cite{nkcd}
\bea
\frac{dE^{QCD}_{\rm flux}(\tau')}{d\tau'}=[\frac{<0| \sum_{r'''}{\hat {\cal Q}}^H(\tau''',r''') \sum_{q,{\bar q},g} \int d^3x' \partial_k  T^{k 0}_{QCD}(\tau',x'){\hat {\cal Q}}^H(0) |0>}{<0| \sum_{r'''}{\hat {\cal Q}}^H(\tau''',r''') {\hat {\cal Q}}^H(0) |0>}]_{\tau'''\rightarrow \infty}.
\label{efd}
\eea
where $T^{\mu \nu}_{QCD}(x)$ is the energy-momentum tensor density in QCD given by \cite{nkntd}
\bea
T^{\mu \sigma}_{QCD}(x)={\bar \psi}_i(x) \gamma^\mu [\delta^{ij}i \partial^\sigma -i gT^a_{ij}A^{a \sigma}(x)]\psi_j(x)+F^{\mu \delta a}(x)F_{\delta }^{~\sigma a}(x)+\frac{g^{\mu \sigma}}{4} F^{\nu \delta a}(x)F_{\nu \delta}^a(x).\nonumber \\
\label{emdd}
\eea
Using eqs. (\ref{hed}) and (\ref{efd}) in (\ref{pi5d}) for the hadron at rest we find
\bea
|<0|{\hat {\cal Q}}^H(0)|H>|^2~e^{-\tau' M_H}=\left[\frac{\sum_{r'} <0|{\hat {\cal Q}}^H(\tau',r') {\hat {\cal Q}}^H(0)|0>}{e^{[\frac{<0| \sum_{r'''}{\hat {\cal Q}}^H(\tau''',r''') \sum_{q,{\bar q},g} \int d\tau' \int d^3x' \partial_k  T^{k 0}_{QCD}(\tau', x'){\hat {\cal Q}}^H(0) |0>}{<0| \sum_{r'''}{\hat {\cal Q}}^H(\tau''',r''') {\hat {\cal Q}}^H(0) |0>}]_{\tau'''\rightarrow \infty}}}\right]_{\tau' \rightarrow \infty}
\label{hadd}
\eea
where $M_H$ is the mass of the hadron, $\int d\tau'$ is the indefinite integration and $\int d^3r'$ is the definite integration.

Eq. (\ref{hadd}) is the correct formula to study the hadron formation from the quarks and gluons by using the lattice QCD method at the zero temperature.

\section{Hadron formation from the thermalized quark-gluon plasma using lattice QCD method at finite temperature}

Since eq. (\ref{hadd}) in the zero temperature QCD is in the Euclidean time one finds it useful to extend the eq. (\ref{hadd}) to the finite temperature QCD because the Euclidean time (imaginary time) can be related to the inverse of the temperature as follows \cite{nkftd}
\bea
-i\int_0^\infty dt \rightarrow \int_0^{\frac{1}{T}}d \tau.
\label{imgd1}
\eea
In order to calculate the medium average in statistical physics using the quantum field theory one finds that the quantum field satisfies the periodic boundary condition, {\it i. e.}, the quark and gluon fields at finite temperature QCD satisfy the periodic boundary conditions
\bea
A_\lambda^d(\tau,r) =A_\lambda^d(\tau+\frac{1}{T}),~~~~~~~~~~~~~\psi_k(\tau,r)=\psi_k(\tau+\frac{1}{T},r).
\label{pbd}
\eea
The non-perturbative correlation function in the finite temperature QCD is given by \cite{nkftd}
\bea
&&<in|{ {\cal Q}}^H(\tau',r') { {\cal Q}}^H(0)|in>=\frac{1}{Z[0]}\int [dA] [d{\bar \psi}][d\psi]{ {\cal Q}}^H(\tau',r') { {\cal Q}}^H(0)\times {\rm det}[\frac{\delta J_f^a}{\delta \omega^b}]\times {\rm exp}[-\int_0^{\frac{1}{T}}d\tau \nonumber \\
&&\int d^3r [-\frac{1}{4}F_{\nu \sigma }^a(\tau,r)F^{\nu \sigma a}(\tau,r)
-\frac{1}{22\alpha} [J_f^d(\tau,r)]^2+{\bar \psi}_{i}(\tau,r)[\delta^{ij}(i{\not \partial} -m)+gT^d_{ij}\aslash^d(\tau,r)]\psi_{j}(\tau,r)]] \nonumber \\
\label{ftqd}
\eea
where $|in>$ is the non-perturbative ground state in the finite temperature QCD.

Using eq. (\ref{ted}) in (\ref{ftqd}) we find
\bea
&&<in| {\cal Q}^H(t',r') {\cal Q}^H(0)|in> = <in| e^{-itH^{QCD}_{q{\bar q}g}}{\cal Q}^H(0,r') e^{it'H^{QCD}_{q{\bar q}g}} {\cal Q}^H(0)|in>.
\label{cffd}
\eea
Hence inserting the complete set of hadronic energy eigenstates as given by eq. (\ref{csd}) in eq. (\ref{cffd}) we find
\bea
&&<in| {\cal Q}^H(t',r') {\cal Q}^H(0)|in> = \sum_l <in|e^{-it'H^{QCD}_{q{\bar q}g}}{\cal Q}^H(0,r')|H_l><H_l| {\cal Q}^H(0)|in>e^{i\int dt'E^{QCD}_{q{\bar q}g,~l}(t')} \nonumber \\
\label{pi3fd}
\eea
Note that unlike eq. (\ref{zrd}) in zero temperature QCD we find
\bea
H^{QCD}_{q{\bar q}g}|in> \neq 0
\label{zrfd}
\eea
in the finite temperature QCD. Using eq. (\ref{zrfd}) in (\ref{pi3fd}) we find
\bea
&&\sum_{r'}<in| {\cal Q}^H(t',r') {\cal Q}^H(0)|in> \neq \sum_l |<in|{\cal Q}^H(0)|H_l>|^2e^{i\int dt'E^{QCD}_{q{\bar q}g,~l}(t')}.
\label{pi4fd}
\eea

Hence from eq. (\ref{pi4fd}) we find that unlike the hadron formation using the non-perturbative correlation function $<0|{\hat {\cal Q}}^H(\tau',r') {\hat {\cal Q}}^H(0)|0>$ in eq. (\ref{hadd}) in QCD in vacuum we cannot use $<in|{\hat {\cal Q}}^H(\tau',r') {\hat {\cal Q}}^H(0)|in>$ to study the hadron formation from the thermalized quark-gluon plasma.

Instead, the non-perturbative correlation function
\bea
&&\sum_{r'}<in| e^{it'H^{QCD}_{q{\bar q}g}}{\cal Q}^H(t',r') {\cal Q}^H(0)|in> = \sum_l |<in|{\cal Q}^H(0)|H_l>|^2e^{i\int dt'E^{QCD}_{q{\bar q}g,~l}(t')}
\label{pi5fda}
\eea
which in the Euclidean time (in the finite temperature QCD formalism) becomes
\bea
&&\sum_{r'}<in| e^{-\tau'H^{QCD}_{q{\bar q}g}}{\cal Q}^H(\tau',r') {\cal Q}^H(0)|in> = \sum_l |<in|{\cal Q}^H(0)|H_l>|^2e^{-\int d\tau'E^{QCD}_{q{\bar q}g,~l}(\tau')}
\label{pi5fd}
\eea
can be used to study the hadron production from thermalized quark-gluon plasma where
\bea
&&<in|e^{-\tau'H^{QCD}_{q{\bar q}g}}{ {\cal Q}}^H(\tau',r') { {\cal Q}}^H(0)|in>=\frac{1}{Z[0]}\int [dA] [d{\bar \psi}][d\psi]e^{-\tau'H^{QCD}_{q{\bar q}g}}{ {\cal Q}}^H(\tau',r') { {\cal Q}}^H(0)\times {\rm det}[\frac{\delta J_f^a}{\delta \omega^b}]\nonumber \\
&&\times {\rm exp}[-\int_0^{\frac{1}{T}}d\tau \int d^3r [-\frac{1}{4}F_{\nu \sigma }^a(\tau,r)F^{\nu \sigma a}(\tau,r)
-\frac{1}{22\alpha} [J_f^d(\tau,r)]^2+{\bar \psi}_{i}(\tau,r)[\delta^{ij}(i{\not \partial} -m)\nonumber \\
&&+gT^d_{ij}\aslash^d(\tau,r)]\psi_{j}(\tau,r)]].
\label{ftqfd}
\eea
Note that the limit of $\tau$ from $0$ to $\frac{1}{T}$ in eq. (\ref{ftqfd}) is necessary to evaluate the thermal average of partons inside quark-gluon plasma which is the standard procedure in the finite temperature quantum field theory to evaluate the thermal average to correspond to the statistical physics. However, the hadron can not be formed inside the quark-gluon plasma because the quark-gluon plasma belongs to the deconfined phase of QCD whereas the hadron belongs to the confined phase of QCD separated by the deconfinement temperature $T_c$. Hence the hadron is formed outside the quark-gluon plasma medium. Since the hadron is formed in vacuum (outside the quark-gluon plasma medium) one finds that the limit of $\tau'$ can go to $ \infty$ in eq. (\ref{ftqfd}) even if the limit of $\tau$ goes to $\frac{1}{T}$.

Hence by neglecting the higher energy level contribution of the hadron at the large Euclidean time [similar to eq. (\ref{pi5d})] we find from eq. (\ref{pi5fd})
\bea
&&[\sum_{r'}<in|e^{-\tau'H^{QCD}_{q{\bar q}g}} {\cal Q}^H(\tau',r') {\cal Q}^H(0)|in>]_{\tau' \rightarrow \infty} = |<in|{\cal Q}^H(0)|H>|^2e^{-\int d\tau'E^{QCD}_{q{\bar q}g}(\tau')}.\nonumber \\
\label{pi5df}
\eea
Since the hadron is formed in the vacuum (outside the quark-gluon plasma medium) the eqs. (\ref{hed}) and (\ref{efd}) are applicable in eq. (\ref{pi5df}).

Hence using eqs. (\ref{hed}) and (\ref{efd}) in (\ref{pi5df}) we find [similar to eq. (\ref{hadd})] for the hadron at rest \cite{nkftd}
\bea
|<in|{\hat {\cal Q}}^H(0)|H>|^2=\left[\frac{\sum_{r'} <in|e^{-\tau'H^{QCD}_{q{\bar q}g}}{\hat {\cal Q}}^H(\tau',r') {\hat {\cal Q}}^H(0)|in>}{e^{[\frac{<0| \sum_{r'''}{\hat {\cal Q}}^H(\tau''',r''') \sum_{q,{\bar q},g} \int d\tau' \int d^3x' \partial_k  T^{k 0}_{QCD}(\tau', x'){\hat {\cal Q}}^H(0) |0>}{<0| \sum_{r'''}{\hat {\cal Q}}^H(\tau''',r''') {\hat {\cal Q}}^H(0) |0>}]_{\tau'''\rightarrow \infty}}}\right]_{\tau' \rightarrow \infty} e^{\tau' M_H}. \nonumber \\
\label{hadfd}
\eea
Hence the hadron formation from thermalized quark-gluon plasma can be studied from eq. (\ref{hadfd}) by using the lattice QCD method at the finite temperature.

\section{Hadron formation from non-equilibrium quark-gluon plasma at RHIC and LHC by using closed-time path integral formalism in non-equilibrium QCD}

Note that in non-equilibrium quantum field theory we cannot use the Euclidean time (imaginary time) as given by eqs. (\ref{imgd}) and (\ref{imgd1}) because the non-equilibrium quantum field theory is formulated in the real time. Hence if we extend the analysis of sections II and III to non-equilibrium QCD then we will not get the exponential decay behavior as given by eqs. (\ref{pi4d}) and (\ref{pi5fd}) but rather we will obtain the oscillating behavior such as $e^{i\int dt'E^{QCD}_{q{\bar q}g,~l}(t')}$. Therefore in non-equilibrium QCD we cannot neglect the higher energy level contributions of the hadron at the large time as we did in eqs. (\ref{pi5d}) and (\ref{pi5df}) for the QCD in vacuum and for the QCD at the finite temperature respectively. This is the main problem to study hadron formation in non-equilibrium QCD which needs to be handled carefully.

In order to study the hadron formation from non-equilibrium quark-gluon plasma by using closed-time path integral formalism in non-equilibrium QCD we proceed as follows.

The non-perturbative correlation function in the closed-time path integral formalism in non-equilibrium QCD is given by \cite{nk1d}
\bea
&&<in|{ {\cal Q}}^H_+(t',r') { {\cal Q}}^H_+(0)|in>=\int [dA_+] [dA_-][d{\bar \psi}_+][d\psi_+][d{\bar \psi}_-][d\psi_-]{ {\cal Q}}^H_+(t',r') { {\cal Q}}^H_+(0)\times {\rm det}[\frac{\delta J_{f+}^a}{\delta \omega^b_+}]\nonumber \\
&& \times {\rm det}[\frac{\delta J_{f-}^a}{\delta \omega^b_-}]\times {\rm exp}[i\int d^4x [-\frac{1}{4}F_{\nu \sigma +}^a(x)F^{\nu \sigma a}_+(x)+\frac{1}{4}F_{\nu \sigma -}^a(x)F^{\nu \sigma a}_-(x)
-\frac{1}{22\alpha} [J_{f+}^d(x)]^2+\frac{1}{22\alpha} [J_{f-}^d(x)]^2\nonumber \\
&&+{\bar \psi}_{i+}(x)[\delta^{ij}(i{\not \partial} -m)+gT^d_{ij}\aslash^d_+(x)]\psi_{j+}(x)-{\bar \psi}_{i-}(x)[\delta^{ij}(i{\not \partial} -m)+gT^d_{ij}\aslash^d_-(x)]\psi_{j-}(x)]] \nonumber \\
&& \times <A_+,{\bar \psi}_+,\psi_+,0|\rho|0,\psi_-,{\bar \psi}_-,A_->
\label{ftqdn}
\eea
where $|in>$ is the non-perturbative ground state in non-equilibrium QCD, $\rho$ is the initial density of states, $\pm$ are closed-time path indices and
\bea
F_{\nu \lambda \pm}^c(x) =\partial_\nu A_{\lambda \pm}^c(x)-\partial_\lambda A_{\nu \pm}^c(x) +gf^{chs} A_{\nu \pm}^h(x) A_{\lambda \pm}^s(x).
\label{fnldn}
\eea

Using eq. (\ref{ted}) in (\ref{ftqdn}) we find
\bea
&&<in| {\cal Q}^H_+(t',r') {\cal Q}^H_+(0)|in> = <in| e^{-itH^{QCD}_{q{\bar q}g+}}{\cal Q}^H_+(0,r') e^{it'H^{QCD}_{q{\bar q}g+}} {\cal Q}^H_+(0)|in>.
\label{cffdn}
\eea
Note that one can also formulate the QCD in vacuum using the closed-time path integral formalism, however it is not necessary as it brings unnecessary complications.
Hence, as mentioned in section III, since the hadron $H$ is formed in vacuum (outside the quark-gluon plasma plasma medium) we find
\bea
H^{QCD}_{q{\bar q}g+}|H_l> = E^{QCD}_{q{\bar q}g,~l}(t)|H_l>
\label{had1}
\eea
where $E^{QCD}_{q{\bar q}g,~l}(t)$ is the energy of all the quarks plus antiquarks plus gluons inside the hadron $H$ in its l$th$ level where the hadron $H$ is formed in vacuum (outside the quark-gluon plasma plasma medium).

Therefore inserting the complete set of hadronic energy eigenstates as given by eq. (\ref{csd}) in eq. (\ref{cffdn}) and then using eq. (\ref{had1}) we find
\bea
&&<in| {\cal Q}^H_+(t',r') {\cal Q}^H_+(0)|in> = \sum_l <in|e^{-it'H^{QCD}_{q{\bar q}g+}}{\cal Q}^H_+(0,r')|H_l><H_l| {\cal Q}^H_+(0)|in>e^{i\int dt'E^{QCD}_{q{\bar q}g,~l}(t')} \nonumber \\
\label{pi3fdn}
\eea
Note that unlike eq. (\ref{zrd}) in zero temperature QCD we find
\bea
H^{QCD}_{q{\bar q}g+}|in> \neq 0
\label{zrfdn}
\eea
in non-equilibrium QCD. Using eq. (\ref{zrfdn}) in (\ref{pi3fdn}) we find
\bea
&&<in| {\cal Q}^H_+(t',r') {\cal Q}^H_+(0)|in> \neq \sum_l <in|{\cal Q}^H_+(0,r')|H_l><H_l| {\cal Q}^H_+(0)|in>e^{i\int dt'E^{QCD}_{q{\bar q}g,~l}(t')}.\nonumber \\
\label{pi4fdn}
\eea

Hence from eq. (\ref{pi4fdn}) we find that unlike the hadron formation using the non-perturbative correlation function $<0|{\hat {\cal Q}}^H(t',r') {\hat {\cal Q}}^H(0)|0>$ in eq. (\ref{hadd}) in QCD in vacuum we cannot use $<in|{\hat {\cal Q}}^H_+(t',r') {\hat {\cal Q}}^H_+(0)|in>$ to study the hadron formation from the non-equilibrium quark-gluon plasma.

Instead, the non-perturbative correlation function
\bea
&&\sum_{r'}<in| e^{it'H^{QCD}_{q{\bar q}g+}}{\cal Q}^H_+(t',r') {\cal Q}^H_+(0)|in> = \sum_l |<in|{\cal Q}^H_+(0)|H_l>|^2e^{i\int dt'E^{QCD}_{q{\bar q}g,~l}(t')}
\label{pi5fdan}
\eea
is non-equilibrium QCD is similar to eq. (\ref{pi5fda}) in finite temperature QCD.

Note that in finite temperature QCD we could go to Euclidean time to find eq. (\ref{pi5fd}) which contains the factor $e^{-\int d\tau'E^{QCD}_{q{\bar q}g,~l}(\tau')}$ which is exponentially falling so that we could neglect the higher energy level contributions of hadron at the large Euclidean time to obtain eq. (\ref{pi5df}).

However, we cannot follow the same procedure in non-equilibrium QCD because we can not go to Euclidean time (imaginary time) in non-equilibrium QCD because the non-equilibrium quantum field theory is formulated in real time (Minkowski time). Therefore we can not obtain the exponentially falling factor from eq. (\ref{pi5fdan}) in non-equilibrium QCD.

Hence we find that, in order to study hadron production from the non-equilibrium quark-gluon plasma, the corresponding non-perturbative partonic correlation function in non-equilibrium QCD using the closed-time path integral formalism is given by
\bea
&&<in|{ {\cal Q}}^H_+(0,r') e^{-t'H^{QCD}_{q{\bar q}g+}}{ {\cal Q}}^H_+(0)|in>=\int [dA_+] [dA_-][d{\bar \psi}_+][d\psi_+][d{\bar \psi}_-][d\psi_-]{ {\cal Q}}^H_+(0,r')e^{-t'H^{QCD}_{q{\bar q}g+}} { {\cal Q}}^H_+(0)\nonumber \\
&&\times {\rm det}[\frac{\delta J_{f+}^a}{\delta \omega^b_+}] \times {\rm det}[\frac{\delta J_{f-}^a}{\delta \omega^b_-}]\times {\rm exp}[i\int d^4x [-\frac{1}{4}F_{\nu \sigma +}^a(x)F^{\nu \sigma a}_+(x)+\frac{1}{4}F_{\nu \sigma -}^a(x)F^{\nu \sigma a}_-(x)
-\frac{1}{22\alpha} [J_{f+}^d(x)]^2\nonumber \\
&&+\frac{1}{22\alpha} [J_{f-}^d(x)]^2+{\bar \psi}_{i+}(x)[\delta^{ij}(i{\not \partial} -m)+gT^d_{ij}\aslash^d_+(x)]\psi_{j+}(x)-{\bar \psi}_{i-}(x)[\delta^{ij}(i{\not \partial} -m)\nonumber \\
&&+gT^d_{ij}\aslash^d_-(x)]\psi_{j-}(x)]] \times <A_+,{\bar \psi}_+,\psi_+,0|\rho|0,\psi_-,{\bar \psi}_-,A_->.
\label{ftqdn0}
\eea
Inserting the complete set of hadronic energy eigenstates as given by eq. (\ref{csd}) in eq. (\ref{ftqdn0}) and then using eq. (\ref{had1}) we find
\bea
&&\sum_{r'}<in| {\cal Q}^H_+(0,r') e^{-t'H^{QCD}_{q{\bar q}g+}}{\cal Q}^H_+(0)|in> = \sum_l |<in|{\cal Q}^H_+(0)|H_l>|^2e^{-\int dt'E^{QCD}_{q{\bar q}g,~l}(t')}
\label{pi5fdan0}
\eea
which contains the corresponding exponentially falling factor $e^{-\int dt'E^{QCD}_{q{\bar q}g,~l}(t')}$.

Hence by neglecting the higher energy level contribution of the hadron at the large time [similar to eqs. (\ref{pi5d}) and (\ref{pi5df}) in QCD in vacuum and in finite temperature QCD respectively] we find in non-equilibrium QCD from eq. (\ref{pi5fdan0})
\bea
&&[\sum_{r'}<in| {\cal Q}^H_+(0,r') e^{-t'H^{QCD}_{q{\bar q}g+}}{\cal Q}^H_+(0)|in>]_{t'\rightarrow \infty} = |<in|{\cal Q}^H_+(0)|H>|^2e^{-\int dt'E^{QCD}_{q{\bar q}g}(t')}.\nonumber \\
\label{pi5df0}
\eea
Since the hadron is formed in the vacuum (outside the quark-gluon plasma medium) the eq. (\ref{hed}) is applicable in eq. (\ref{pi5df0}). Hence using eq. (\ref{hed}) in (\ref{pi5df0}) we find
\bea
&&[\sum_{r'}<in| {\cal Q}^H_+(0,r') e^{-t'H^{QCD}_{q{\bar q}g+}}{\cal Q}^H_+(0)|in>]_{t'\rightarrow \infty} = |<in|{\cal Q}^H_+(0)|H>|^2e^{-t'E^H}e^{\int dt'E^{QCD}_{{\rm flux}}(t')}.\nonumber \\
\label{pi5df01}
\eea
Note that the eq. (\ref{efd}) for $E^{QCD}_{{\rm flux}}(\tau')$ is in Euclidean time in QCD in vacuum which cannot be used in eq. (\ref{pi5df01}) because the eq. (\ref{pi5df01}) is in real time. Extending eq. (\ref{efd}) to real time (Minkowski time) in QCD in vacuum [similar to eq. (\ref{ftqdn0}) in non-equilibrium QCD] we find
\bea
\frac{dE^{QCD}_{\rm flux }(t')}{dt'}=[\frac{<0| \sum_{r'''}{\hat {\cal Q}}^H(0,r''')e^{-t'''H^{QCD}_{q{\bar q}g}} \sum_{q,{\bar q},g} \int d^3x' \partial_k  T^{k 0}_{QCD}(t',x'){\hat {\cal Q}}^H(0) |0>}{<0| \sum_{r'''}{\hat {\cal Q}}^H(0,r''')e^{-t'''H^{QCD}_{q{\bar q}g}} {\hat {\cal Q}}^H(0) |0>}]_{t'''\rightarrow \infty}.\nonumber \\
\label{efdn}
\eea

Using eq. (\ref{efdn}) in (\ref{pi5df01}) we find [similar to eqs. (\ref{hadd}) and (\ref{hadfd})] for the hadron at rest \cite{nkftd}
\bea
|<in|{\hat {\cal Q}}^H_+(0)|H>|^2=\left[\frac{\sum_{r'} <in|{\hat {\cal Q}}^H_+(0,r') e^{-t'H^{QCD}_{q{\bar q}g}}{\hat {\cal Q}}^H_+(0)|in>}{e^{[\frac{<0| \sum_{r'''}{\hat {\cal Q}}^H(0,r''') e^{-t'''H^{QCD}_{q{\bar q}g}}\sum_{q,{\bar q},g} \int dt' \int d^3x' \partial_k  T^{k 0}_{QCD}(t', x'){\hat {\cal Q}}^H(0) |0>}{<0| \sum_{r'''}{\hat {\cal Q}}^H(0,r''')e^{-t'''H^{QCD}_{q{\bar q}g}} {\hat {\cal Q}}^H(0) |0>}]_{t'''\rightarrow \infty}}}\right]_{t' \rightarrow \infty} e^{t' M_H}. \nonumber \\
\label{hadfd0}
\eea

Hence the hadron formation from the non-equilibrium quark-gluon plasma can be studied from eq. (\ref{hadfd0}) by using the closed-time path integral formalism in non-equilibrium QCD.
\section{Conclusions}
Recently we have reported the correct formulation of the hadron formation from the quarks and gluons by using the lattice QCD method at the zero temperature. Similarly we have also reported the correct formulation of the hadron formation from the thermalized quark-gluon plasma by using the lattice QCD method at the finite temperature. In this paper we have extended this to non-equilibrium QCD and have presented the correct formulation of the hadron formation from the non-equilibrium quark-gluon plasma by using the closed time path integral formalism. Hadron formation from the non-equilibrium quark-gluon plasma is necessary to detect the quark-gluon plasma at RHIC and LHC.

\end{document}